\documentclass[conference]{IEEEtran}

\usepackage{amsmath, amssymb}
\usepackage{subfig}
\usepackage{float}
\usepackage{graphicx, float}
\usepackage{multirow}
\usepackage{array}

\usepackage[noadjust]{cite}

\usepackage{ragged2e}
\usepackage{float}
\usepackage{graphicx}
\usepackage{calc}

\usepackage[english]{babel}
\usepackage[utf8x]{inputenc}
\usepackage{xcolor}
\usepackage{color}
\usepackage{caption}
\usepackage{float}
\usepackage{hyperref}
\usepackage{amsmath}
\usepackage{color,soul}

\def\BibTeX{{\rm B\kern-.05em{\sc i\kern-.025em b}\kern-.08em
    T\kern-.1667em\lower.7ex\hbox{E}\kern-.125emX}}
    
\usepackage{fancyhdr} 
\usepackage{kantlipsum} 
\fancypagestyle{plain}{ 
\fancyhf{} 
\fancyhead[C]{Conference on \LaTeX} 
 
} 
\usepackage{eso-pic}

\newcommand\Tstrut{\rule{0pt}{2.6ex}}         

\makeatletter
\newcommand*{\rom}[1]{\expandafter\@slowromancap\romannumeral #1@}
\makeatother

\makeatletter
\begingroup\lccode`~=`\%
\lowercase{\endgroup\def~{\new@ifnextchar~\tohecz@comment\%}}
\def\tohecz@comment{\catcode`\^^M=3 \tohecz@commentignore}
\begingroup\lccode`$=`\^^M
\lowercase{\endgroup\def\tohecz@commentignore#1$}{\catcode`\^^M=5 }
\makeatother

\usepackage{todonotes}

\urldef{\mails}\path|{firstname.lastname}@uni-passau.de|

\begin{document}

%
%
%
%

\title{Predicting Event Attendance exploring Social Influence }

\author{%
  \IEEEauthorblockN{%
    Fatemeh Salehi Rizi and Michael Granitzer
  }
  \IEEEauthorblockA{%
    Department of Computer Science and Mathematics\\
    University of Passau, Germany\\
    Email: \mails
  }
}

\maketitle

\begin{abstract}
The problem of predicting people's participation in real-world
events has received considerable attention as it
offers valuable insights for human behavior analysis and
event-related advertisement. Today social networks (e.g. Twitter) widely reflect large popular events  
 where people discuss their
interest with friends. Event participants usually stimulate friends to join the event which propagates a social influence in the network. In this paper, we propose to model the social influence of friends on event attendance.
We consider non-geotagged posts besides structures of social groups to infer users' attendance. 
To leverage the information on network topology we apply some of recent graph embedding techniques such as node2vec, HARP and Poincar\`e.
We describe the
approach followed to design the feature space and feed it to a neural network.
The performance evaluation is conducted using two large music festivals datasets,
namely the VFestival and Creamfields.
The experimental results show that our classifier outperforms the state-of-the-art
baseline with 89$\%$ accuracy observed for the VFestival dataset.

\end{abstract}

%


\maketitle

\section{Introduction}
\label{sec:intro}

Social media platforms such as Twitter provide a great opportunity for people
to communicate about real-world events. Large popular events such as music festivals
attract thousands of participants who usually post their feelings or opinions in a social media. Moreover, people often attend to events along with their friends who are already connected in a social media. Event participants can propagate a social influence in the network which affects friends' attendance decisions.
Therefore, analysis of participants' posts besides their network structures can detect the actual attendance to benefit several applications like advertising or mobility management.

Predicting attendance exploiting geotagged posts is a simple task since the user position 
is associated to the event location.
However, few users indeed make the geotagging~\cite{leetaru2013mapping} during the event which yields a sparse dataset. Inferring attendance would be possible for the few users who shared the location.  

Previous research has studied that people usually participate in events with their friends or family members who have linked in a social network~\cite{zhang2017event,du2014predicting}.
Therefore, users propagate an influence within small social groups which should be considered to infer the attendance. 

We speculate that modeling the social influence by encoding
the structure around a node in the graph improves the prediction task. 
To extract network structural features we apply some of recent graph embedding techniques such as node2vec~\cite{grover2016node2vec}, HARP~\cite{chen2017harp} and Poincar\`e~\cite{Nickel2017} which have shown prominent performance for node classification and link prediction. To understand users' attitudes toward the event, we gather users' posts before and during the event. The posts shared before the event may indicate
user interest and his intention to attend. Posts during the event may express people's feeling by sharing photos and videos during the event.
 



We propose to train a neural network classifier fed by textual features of posts and network topological features. Therefore, the contributions of this paper are as follows.


\begin{itemize}



\item We exploit network topological features applying graph embedding techniques to model users' social influences on event attendance (section \ref{sec:app}).

\item  We inspect the impact of different embedding techniques on the performance of the prediction (section~\ref{sec:exp}).

\item We show our approach outperforms the most prominent work in the state-of-the-art for event attendance prediction (section~\ref{sec:exp}).

\end{itemize}

We conduct our experiments on two real-world datasets from music festivals events held in UK.
The evaluation results show that considering neighborhood structures around users improve the performance as
we achieve accuracy ranging from $88\%$ to
$89\%$.

The rest of the paper is organized as follows. Section~\ref{sec:background}
quickly reviews the related works. In Section~\ref{sec:app}, we describe our
approach for event attendance prediction. Section~\ref{sec:exp} presents
the evaluation results of the predicting approach. The last
section concludes the paper.

\section{Related work}
\label{sec:background}

Twitter is a common platform to reflect events where users post their opinions and feeling. Lira at al.~\cite{de2017exploring} extract the textual and temporal features of tweets to infer users' attendance, however they ignore the impact of friends on users' decisions.
Several research explore Twitter geotagged posts to decide about users' attendance to real-world events. Magnuson et al.~\cite{magnuson2015event} analyze the tweets with geo-location tags to create a recommendation system which offers events based on user's interest, time and geo-location. Similarly~\cite{botta2015quantifying}
investigates whether mobile phone usage and the geo-located
Twitter data can be used to estimate the number of people in
a specific area at a given time. However, we have
a different goal as we do not want to estimate the user's
exact location at the time of the post, but classify the single
posts on the basis of the user's current and past attendance to a given event.

In addition to location based services, event-based social networks (EBSN)
provide a new venue to analyze human social behaviors. Authors in~\cite{zhang2015will} explore users' past activities in EBSN considering the temporal and spatial
factors to predict users' decision on event attendance. But we do not exploit users' history as we aim
at classifying single posts by neglect of the user profiles and specific event information.
There has been some works aim to model social influenec in EBSNs such as \cite{zhang2017event}, ~\cite{liu2012event} and ~\cite{du2014predicting}. However, we do not specifically deal with ESBN graph but instead
focus on popular social networks (e.g. Twitter) where events have been reflected.

There has been numerous techniques to encode topology of a social graph in state-of-the-art.
The most popular works which preserve neighborhood patterns effectively and efficiently are Node2vec \cite{grover2016node2vec}, HARP\cite{chen2017harp} and Poincar\`e \cite{Nickel2017}. 
Node2vec
is a random walk based approach which preserves higher order
proximity between nodes by maximizing the probability of occurrence of subsequent nodes in fixed length random walks. Since large social graphs often form innate hierarchical structures \cite{ravasz2003hierarchical},
we apply Poincar\`e  and HARP to capture hierarchical patterns. Poincar\`e use a hyperbolic space to alleviate overfitting issues that euclidean embeddings face if the
data has intrinsic hierarchical structure. HARP captures
first and second order proximities of neighborhoods by collapsing
the graph into smaller hierarchies.
To our best knowledge, inferring users' attendance to large events exploring the social influence has not been investigated yet.

\section{Problem Formalization}
\label{sec:app}

Given a user $u$ and a future
event $e$, the goal is to predict whether $u$ will attend $e$ or not.
We exploit the
following information to predict the user actual attendance to the event $e$:

\begin{itemize}
\item The post of the user $u$ contains texts and emoticons 
\item The neighborhood structure around the user $u$ in $G$
\end{itemize}

Therefore, we deal with two types of features: \textit{Textual} features and \textit{Network} features.

\textit{Textual} features model the textual content of the post. We
used a Bag of Words (BoW) model with unigrams, bigrams
and trigrams occurring in the post. We apply lemmatization to
group together the different inflected forms of a word. Thus each lemma and each sequence of two and three adjacent lemmas are considered as features. The generated vector is denoted by $T(u)$.

\textit{Network} features model neighborhood patterns around users in a social graph.
Let $G = (V, E)$ be an unweighted graph where the user $u$ and his friends have joined. 
Graph embedding techniques
create vector embedding $ \phi (u) \in R^d$ for every node $u \in V$ such that $ d<<|V|$ . In this work we apply node2vec~\cite{grover2016node2vec}, HARP~\cite{chen2017harp} and Poincar\`e~\cite{Nickel2017} to learn node embeddings.

We propose to use a feedforward neural network to build a binary classifier. Formally, we define $F$ as function,
\[ F: T(u) \oplus \phi(u) \mapsto \{0, 1\} \]
that maps concatenation ($\oplus$) of two feature vectors into a binary value of $1$ for attendance or $0$ for non-attendance.

Our feedforward network consists of an input layer, a hidden layer,
and an output layer. The size of the input layer depends on the size of vector embeddings and tweets vocabulary. 
We set the rectified unit (ReLU)~\cite{nair2010rectified} as activation function of the input and hidden layer. ReLU does not face gradient vanishing problem and it has been shown that deep networks are trained efficiently. 
The number of neurons in the hidden layer is the mean of the neurons in the input and output layers \cite{heaton2008introduction}.
Since the network performs a binary classification task, the output layer is a single unit of Sigmoid~\cite{karlik2011performance}.
As optimizer, we use Adam~\cite{kingma2014adam} which is robust and well-suited to a wide range
of non-convex optimization problems for large-scale learning.


\section{Experimental Evaluation}
\label{sec:exp}

\begin{figure*}[t]
  \centering
  
  \subfloat[VFestival]{\includegraphics[width=0.5\textwidth]{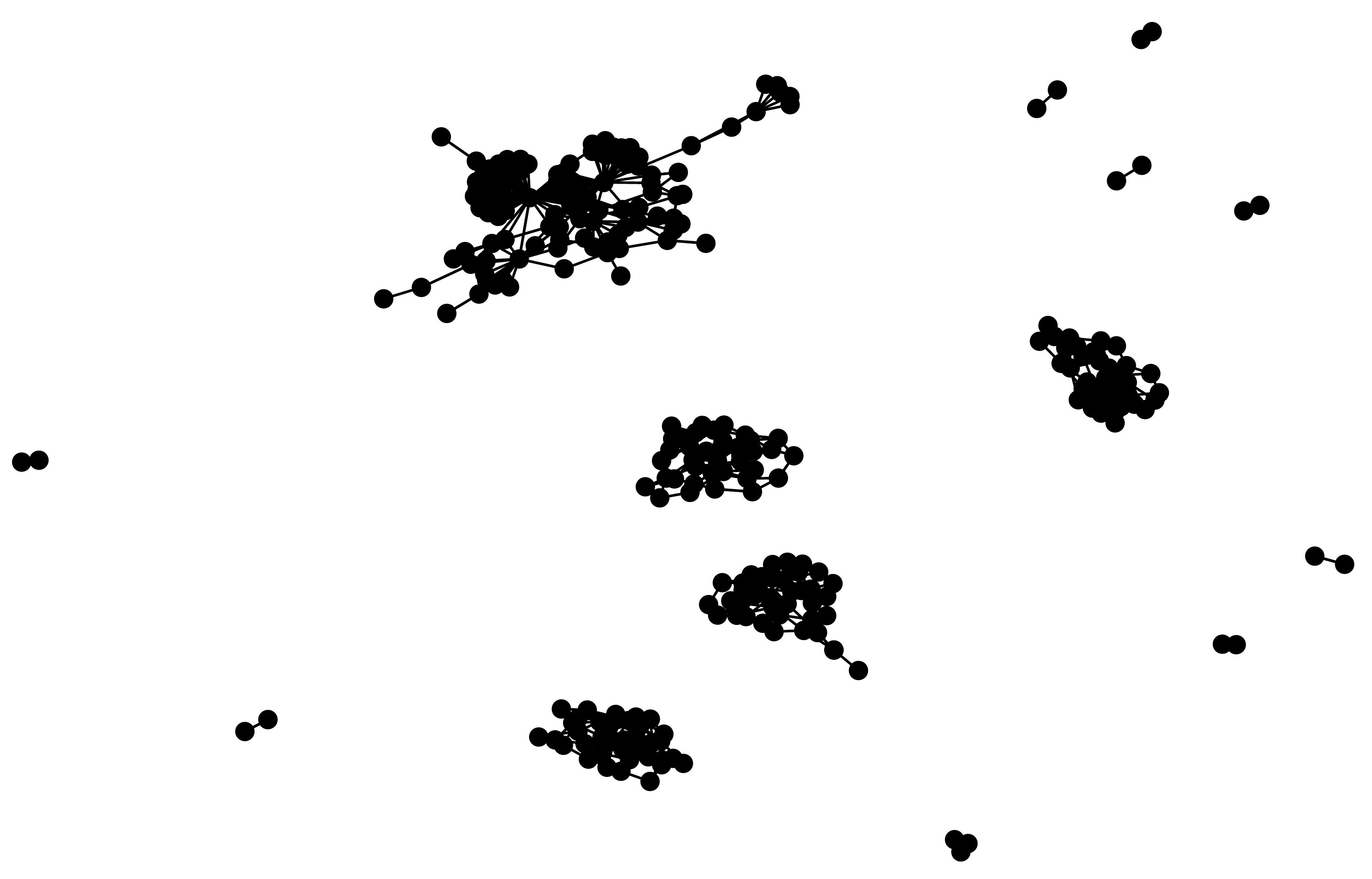}}
  \subfloat[Creamfields]{\includegraphics[width=0.5\textwidth]{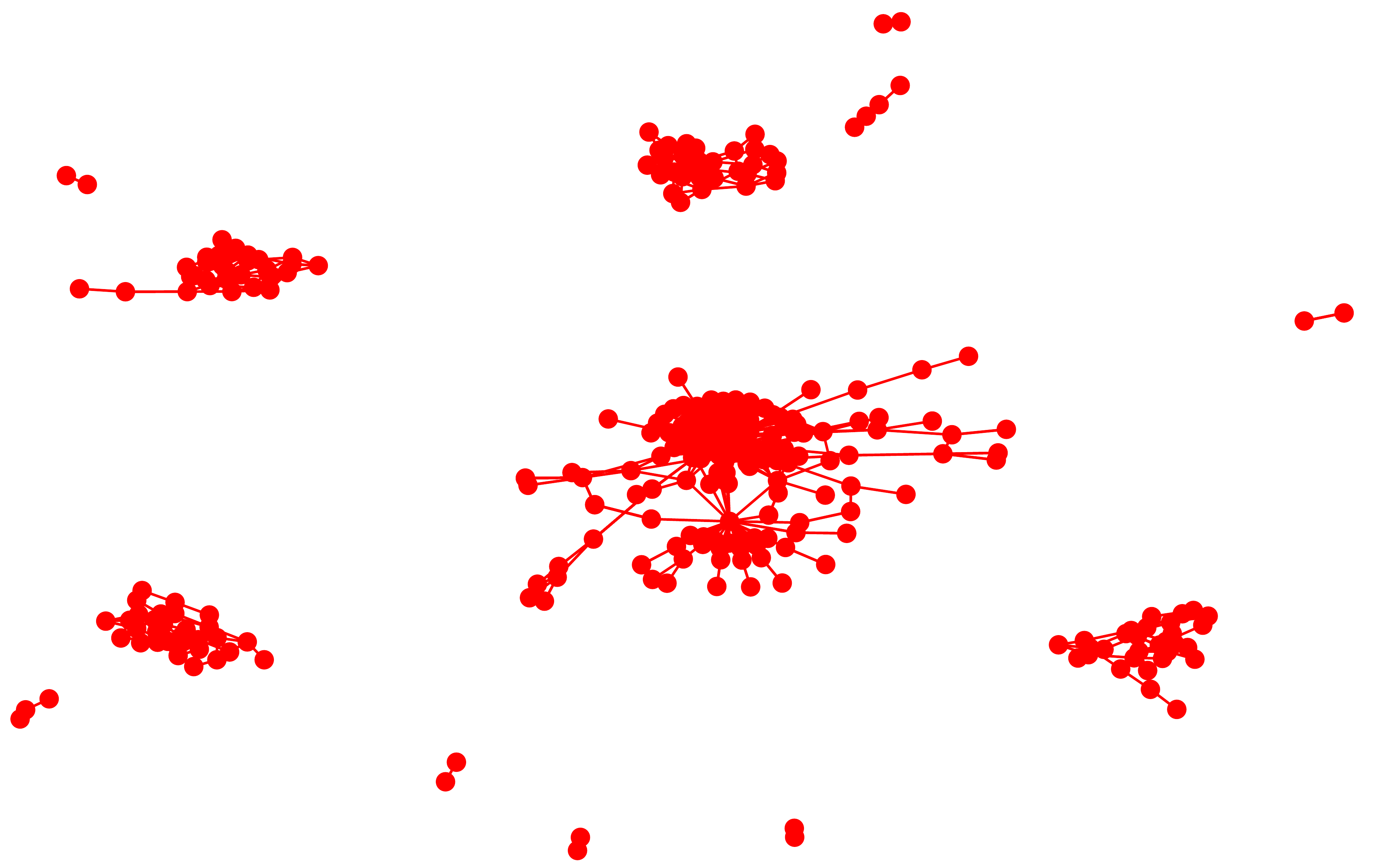}}
  
  \caption{Social groups attended to the events}
  \label{fig:groups}
\end{figure*}
\label{subsec:dataset}
We conduct our experiments on Twitter datasets related to two large music festivals in the UK named Creamfields and VFestival ~\cite{de2017exploring}. The original dataset splits posts based on their timestamp into three different disjoint sets: posts
made before, during or after the event. In each timestamp, $460$ distinct tweets are sampled and labeled. However, we crawled fewer distinct tweets for each timestamp since users deleted their Twitter account after a year. Obviously, we do not have enough training samples if we only consider tweets before the event. Therefore,
we merge the tweets before and during the events to have adequate training samples. 
Table~\ref{tab:dataset} reports the total number of labeled tweets and the
respective percentage of positive (attended) and negative (not-attended) labeled classes for each dataset.

\begin{figure*}[t]
  \centering
  
  \subfloat[HARP]{\includegraphics[width=0.28\textwidth]{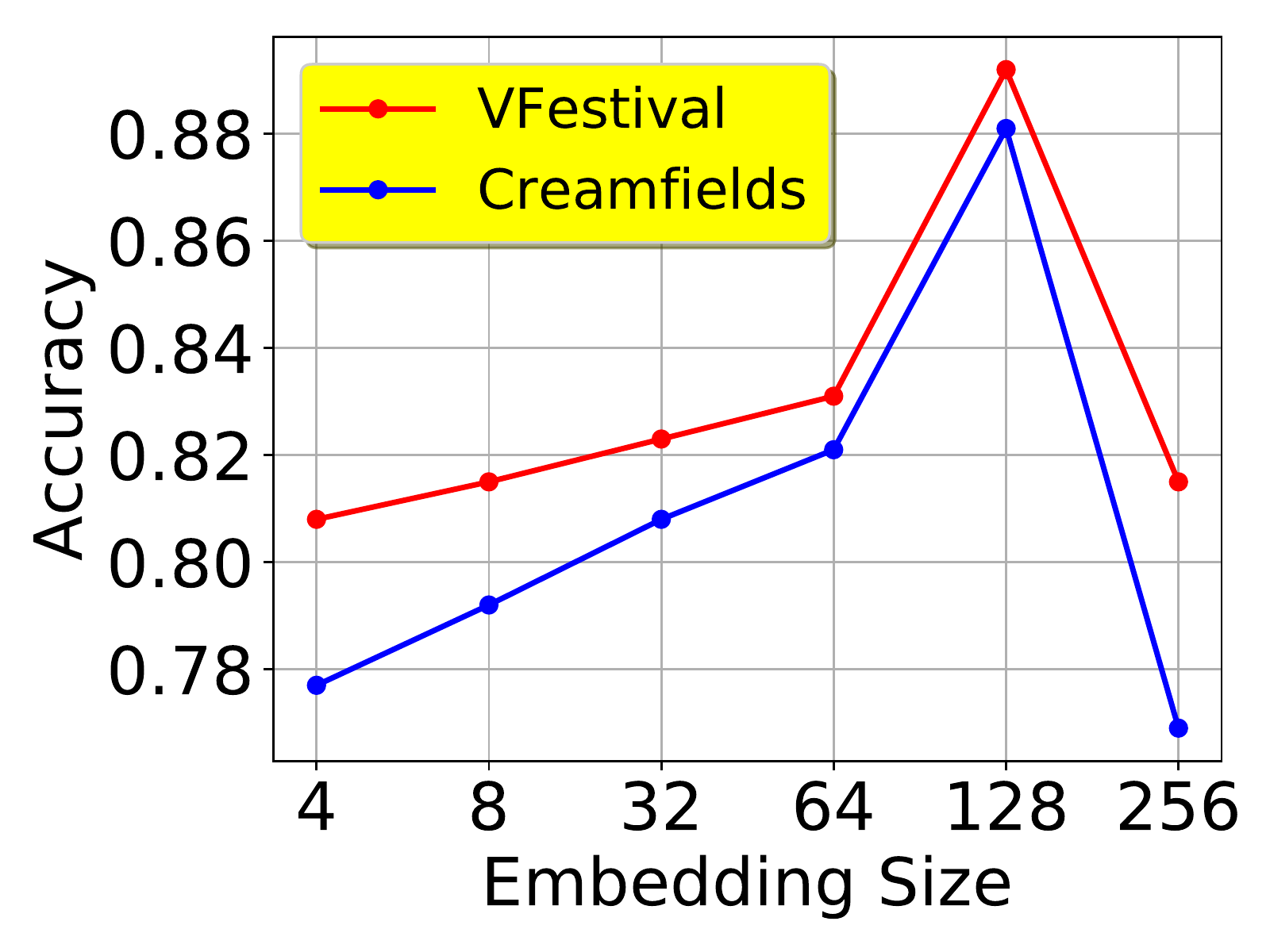}}
  \subfloat[HARP]{\includegraphics[width=0.28\textwidth]{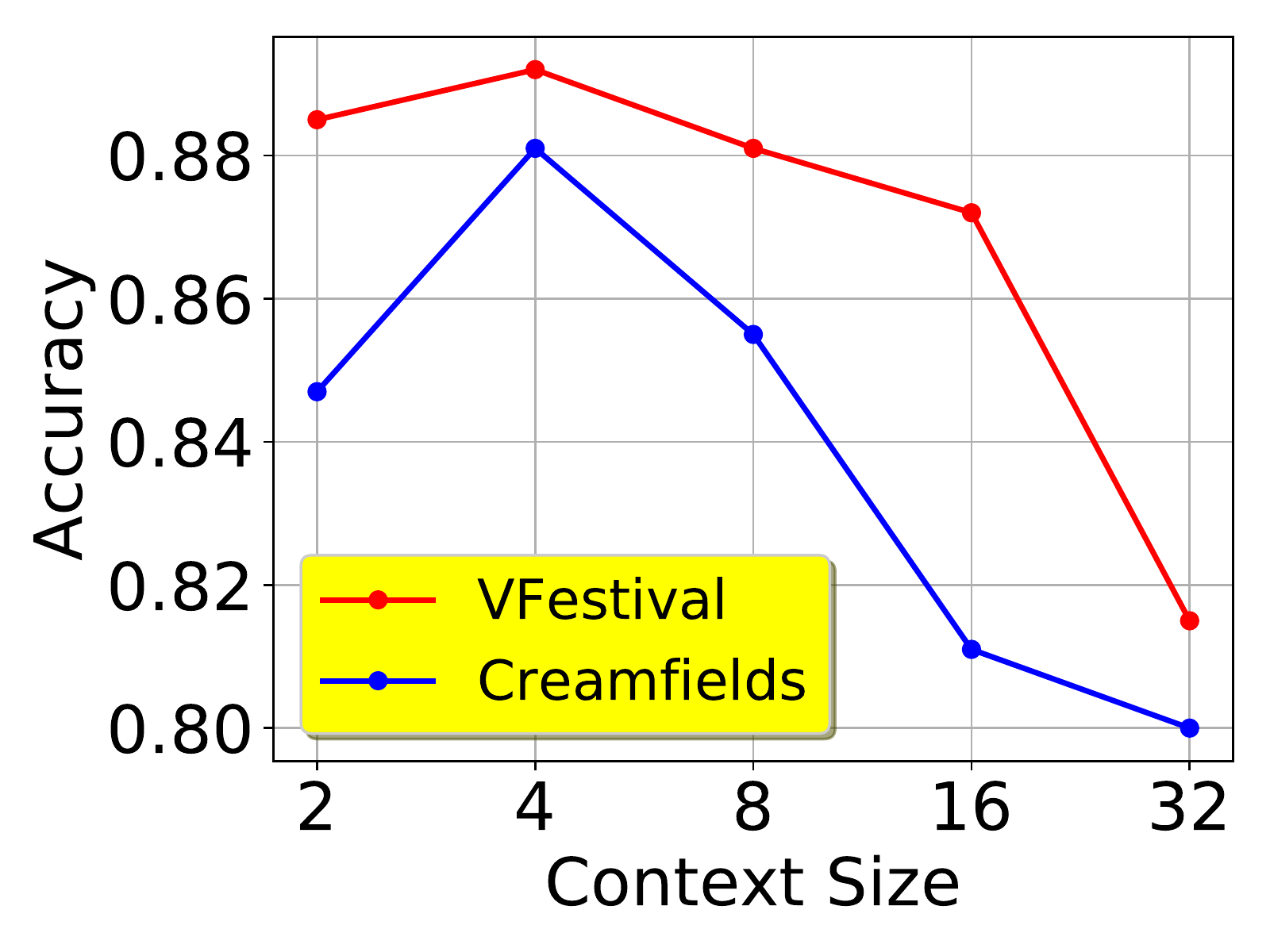}}
  \subfloat[Poincar\'{e}]{\includegraphics[width=0.28\textwidth]{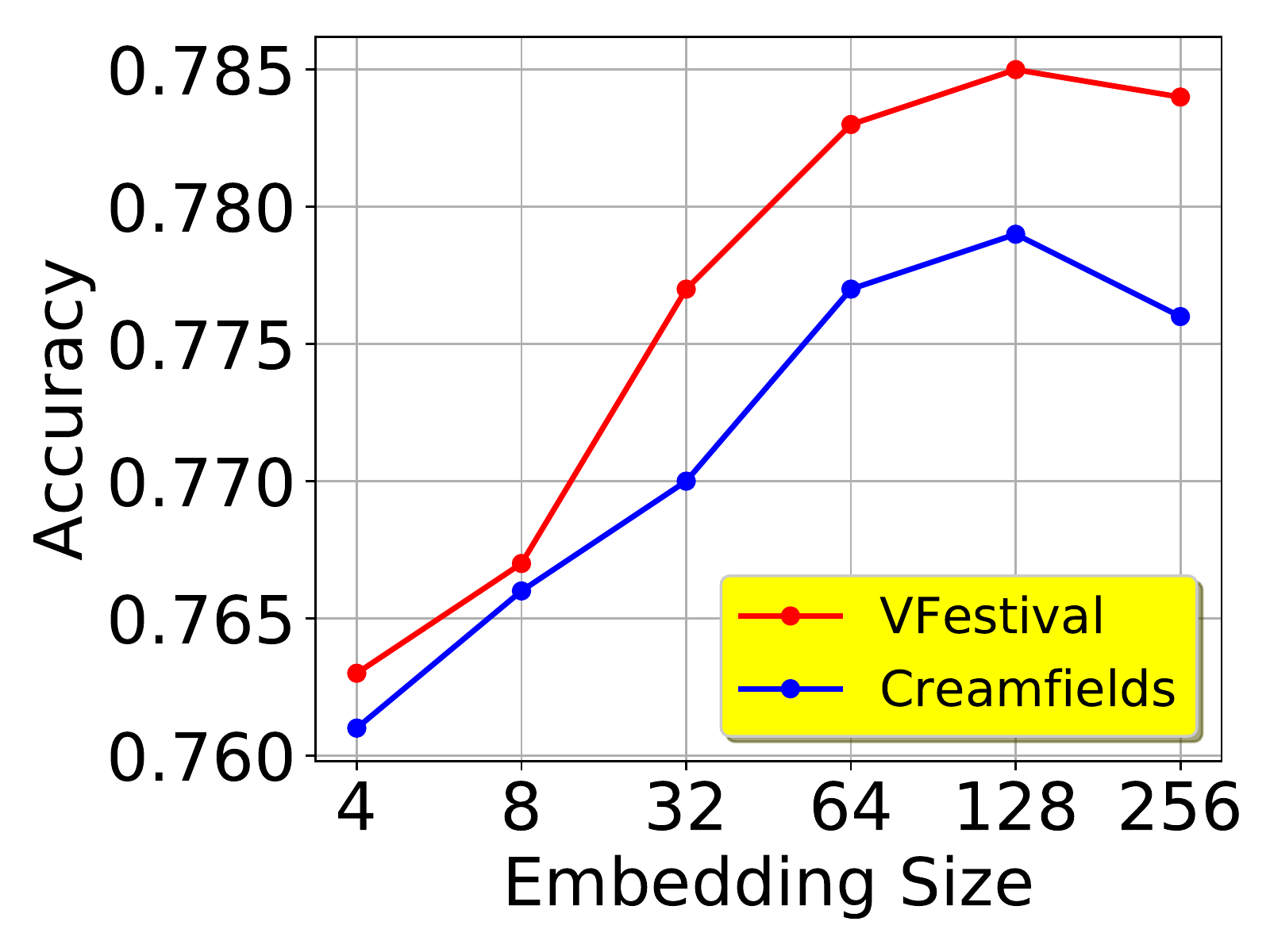}}
  
  \caption{Parameter sensitivity for classification}
  \label{fig:para}
\end{figure*}

\begin{table}[h]
\caption{Dataset Statistics }

\begin{center}
\begin{tabular}{|c|c|c|c|c|}

\hline
\textbf{Dataset}&
\textbf{Timestamp}&
\textbf{Labled}&
\textbf{pos$\%$}&
\textbf{neg$\%$} \Tstrut \\
  
\hline

  \multirow{3}{*}{VFestival} & before  & 320 & 44 & 56 \Tstrut \\ 
                             & during  & 329  & 47  &  53 \Tstrut \\ 
                             & overall & 649 & 46 & 54 \Tstrut  \\

\hline

 \multirow{3}{*}{Creamfields} & before  & 330 & 46 & 54 \Tstrut \\ 
                              & during  & 341 & 42  & 58 \Tstrut \\ 
                              & overall & 671  & 44  & 56 \Tstrut \\
\hline

\end{tabular}
\label{tab:dataset}
\end{center}
\end{table}


To study people's social influence, we use the crawled Twitter graph consists of users and links. Given the list of attended users in the original dataset, we check whether there is an edge between any pair. We thus construct an edge list of attended users in each dataset. There are multiple users participate in the events alone (without direct neighbors). We need to organize alone users into several artificial groups in order to examine friends' influence.
Therefore, we first divide users into 4 individual groups and then randomly link members together. Overall, $303$ users in $14$ groups attended to VFestival and $298$ users participated in Creamfields.
Fig.~\ref{fig:groups} depicts all social groups joined to VFestival and Creamfields. Finally, we apply node2vec, HARP and Poincar\`e to encode neighborhood structures of the social groups in the form of vectors. 

%
%
%

\begin{table*}[t]
\caption{Classification performance applying different graph embedding techniques.}

\begin{center}
\begin{tabular}{|c|l|c|c|c|c|}

\hline
\textbf{Dataset}&
\textbf{Classifier}&
\textbf{Accuracy} &
\textbf{Precision} &
\textbf{Recall} &
\textbf{F1}\Tstrut \\

\hline

  \multirow{6}{*}{VFestival} & GBDT  & 0.805  &0.826  &0.675  & 0.736  \Tstrut\\ 
                             & NN\textsubscript{T$\oplus$HARP} & \textbf{0.892} &\textbf{0.886} & \textbf{0.784}&\textbf{0.821} \Tstrut \\ 
                             & NN\textsubscript{T$\oplus$N2V} & 0.823 & 0.792&0.745 & 0.768  \Tstrut\\ 
                            
                             & NN\textsubscript{T$\oplus$Poincar\`e} & 0.785 & 0.735 & 0.706&  0.720 \Tstrut \\ 
                             
                             & NN\textsubscript{T} & 0.782 & 0.756& 0.667 & 0.708 \Tstrut \\ 
                             & NN\textsubscript{HARP} & 0.769 & 0.691 & 0.745 & 0.717 \Tstrut\\

\hline

 \multirow{6}{*}{Creamfields} &  GBDT & 0.865 &0.838 &0.764 &0.793  \Tstrut\\ 
                              &  NN\textsubscript{T$\oplus$HARP} & \textbf{0.881} & \textbf{0.857} &\textbf{0.774} & \textbf{0.814}  \Tstrut\\
                              & NN\textsubscript{T$\oplus$N2V}&  0.803 & 0.786& 0.758 & 0.752 \Tstrut \\
                              & NN\textsubscript{T$\oplus$Poincar\`e} & 0.779 & 0.735 & 0.689 & 0.712\Tstrut \\ 
                              & NN\textsubscript{T} & 0.777 & 0.729 & 0.686&  0.707   \Tstrut \\
                             & NN\textsubscript{HARP} &0.710 & 0.732 & 0.661 & 0.695 \Tstrut \\
\hline

\end{tabular}
\label{tab:acc}
\end{center}
\end{table*}

\subsection{Parameters and Environment}
\label{subsec:param}

In this section, we empirically study parameter sensitivity of the embedding techniques. 
We examine how the different choices of parameters
affect the performance where the input of the classifier is concatenation of textual and network features. In Fig.~\ref{fig:para}, the accuracy is measured as a function of parameters embedding size $d$ and context size $k$. It can be seen that the performance varies with embedding and context size. In HARP settings, $d=128$ and $k=4$ gives the best performance. 


Since HARP applies a hierarchical paradigm for graph
embedding based on iterative learning methods (e.g. node2vec), we follow the same tuning procedure for node2vec. We also consider $p=q=1$ to explore local and global neighborhood equally. For Poincar\'{e}, we determine the radius $r>1$ and the steepness $t=0.01$ \cite{Nickel2017}. As shown in Fig.~\ref{fig:para}, $d=128$ is the best embedding size for Poincar\'{e} as well.


\subsection{Results and Discussions}
\label{subsec:discu}

Our experiments are conducted using a 4-fold cross validation, while preserving the proportion of positive and negative instances in each fold. The quality of predictor is assessed in terms of accuracy, precision, recall and F1 score. 
We compare our results to the most recent work~\cite{de2017exploring} which achieved the best accuracy using Boosting Decision Trees (GBDT). To make results comparable, we take the average of accuracies before and during the events available in~\cite{de2017exploring}.
Depending on input features to the Neural Network (NN), we have the following classifiers:

\begin{itemize}
\item \textbf{NN\textsubscript{T$\oplus$HARP}}: Input of tweets textual features and HARP embeddings
\item \textbf{NN\textsubscript{T$\oplus$N2V}}: Input of tweets textual features and node2vec embeddings 

\item \textbf{NN\textsubscript{T$\oplus$Poincar\`e}}: Input of tweets textual features and Poincar\`e embeddings 
\item \textbf{NN\textsubscript{T}}: Input of tweets textual features
\item \textbf{NN\textsubscript{HARP}}: Input of HARP embeddings 

\end{itemize}

The obtained accuracies for each dataset are provided in the Table~\ref{tab:acc}. The results indicate that considering the social influence of friends improves the accuracy. We can see that HARP embeddings generally shows the better performance compared to the others. Feeding the network only with HARP embedding which benefits from graph coarsening achieves accuracy ranging from $71\%$ to $76\%$ in two datasets. 
It demonstrates the impact of the local neighborhood structure and friends' influences on the prediction task.
Poincar\`e performs worse than two other embedding techniques. The loss function of Poincar\`e causes most points to
map on the border of the Poincare ball~\cite{ganea2018hyperbolic}. It may cause Poincar\`e shows different performances for preserving local properties~\cite{Bonner2018} which are needed in our prediction task.

\section{Conclusion }
\label{sec:con}
In this paper, we proposed an approach to predict
event attendance by mining users media posts and social groups influence. 
A key detail of our
proposed approach is to exploit user's neighborhood pattern in the social graph to improve the prediction task.
We trained
a neural network classifier using tweets and the social graph embeddings related to two large
music festivals in the UK.
We evaluated the performance by accuracy
and precision in comparison to the recent work.
The results show how our approach performs consistently better
than the baseline, exhibiting $89\%$ accuracy at classifying
attending users.

\section*{Acknowledgment}

The presented work was developed within the Provenance Analytics project funded by the German Federal Ministry of Education and Research, grant agreement number 03PSIPT5C.

\bibliographystyle{IEEEtran}
\bibliography{ref.bib}

\end{document}